\documentclass[english,aps,floats,twocolumn,showpacs,nofootinbib]{revtex4}
\usepackage{pslatex}
\usepackage[T1]{fontenc}
\usepackage[latin1]{inputenc}
\usepackage{graphicx}
\usepackage{epsfig}

\usepackage{calc}
\usepackage{ifthen}

{\newcommand{\lsim}{\mbox{\raisebox{-.6ex}{~$\stackrel{<}{\sim}$~}}}
{\newcommand{\gsim}{\mbox{\raisebox{-.6ex}{~$\stackrel{>}{\sim}$~}}} 
{
\newcommand{\bea}{\begin{eqnarray}}
\newcommand{\eea}{\end{eqnarray}}

\newcommand{\nc}{\newcommand}
\nc{\renc}{\renewcommand}
\nc{\eqs}[2]{\mbox{Eqs.~(\ref{#1},\,\ref{#2})}}
\nc{\eq}[1]{\mbox{Eq.~(\ref{#1})}}
\nc{\figs}[2]{\mbox{Figs.~(\ref{#1},\,\ref{#2})}}
\nc{\fig}[1]{\mbox{Fig~.(\ref{#1})}}
\nc{\be}[1]{\begin{equation} \mbox{$\label{#1}$}}
\nc{\ee}{\vspace{0.1cm}\end{equation}}
\newcommand{\ochi}{\overline{\chi}}

\newcommand{\bean}{\begin{eqnarray*}}
\newcommand{\eean}{\end{eqnarray*}}

\def\GeV{{\rm \ GeV}}

\def\keV{{\rm \ keV}}
\def\TeV{{\rm \ TeV}}
\def\lae{\;^{<}_{\sim} \;} \def\gae{\; ^{>}_{\sim} \;} 

\begin{document}
\title{keV Warm Dark Matter via the Supersymmetric Higgs Portal}
\author{John McDonald}
\email{j.mcdonald@lancaster.ac.uk}
\author{Narendra Sahu}
\email{n.sahu@lancaster.ac.uk}
\affiliation{Cosmology and Astroparticle Physics Group, University of 
Lancaster, Lancaster LA1 4YB, UK}
\begin{abstract}

Warm dark matter (WDM) may resolve the possible conflict between observed galaxy halos and 
the halos produced in cold dark matter (CDM) simulations. Here we present 
an extension of MSSM to include WDM by adding a gauge singlet fermion, $\overline{\chi}$, 
with a portal-like coupling to the MSSM Higgs doublets. 
This model has the property that the dark matter is {\it necessarily warm}.
In the case where $M_{\overline{\chi}}$ 
is mainly due to electroweak symmetry breaking, the $\overline{\chi}$ mass is completely 
determined by its relic density and the reheating temperature, $T_R$.  
For $10^2 \GeV \; 
^{<}_{\sim} \; T_{R} \; ^{<}_{\sim} \; 10^{5} \GeV$, the range allowed by $\ochi$ 
production via thermal Higgs annihilation, the $\overline{\chi}$ mass 
is in the range 0.3-4 keV, precisely the range required for WDM. 
The primordial phase-space density, $Q$, can directly account for that observed in dwarf spheroidal galaxies, $Q 
\approx 5 \times 10^{6} {\rm (eV/cm^3)/(km/s)^3}$, when the reheating temperature is in the range $T_R \approx 10-100$ 
TeV, in which case $M_{\overline{\chi}} \approx 0.45$ keV. The free-streaming 
length is in the range 0.3-4 Mpc, which can be small enough to alleviate the problems of overproduction of galaxy substructure
and low angular momentum of CDM simulations. 

\end{abstract}
\maketitle

{\bf Introduction:} 
Cold dark matter (CDM) with a cosmological constant ($\Lambda$ CDM) is remarkably 
successful in explaining the large scale structure of the observed universe. 
Numerical simulations based on the $\Lambda$CDM model predict cusped central 
densities~\cite{cusp_problem}. The observational situation is less clear. It has been suggested that observed galaxy halos have cores \cite{salucci,gilmore}, implying that CDM has a cusp problem. However, the observational evidence has also been interpreted to support cusped halos \cite{simon}. Here we will consider the former possibility. One way 
to solve the cusp problem, should it exist, is to have dark matter with a sufficiently large velocity 
dispersion, known as warm dark matter (WDM)~\cite{wdm}. In addition, CDM simulations result in overproduction of galactic substructure \cite{satellite_problem} and low angular momentum \cite{angular_momentum}, which may be alleviated by reducing power at small scales via free-streaming of WDM. 

  Several candidates have been suggested to account for WDM, notably sterile neutrinos in 
a minimal extension of the Standard Model (MSM) \cite{dodelson,msm}, and superWIMPs \cite{superwimps}. These models may also account for other phenomena, such as pulsar velocities and baryogenesis in the case of sterile neutrinos, and can have distinctive collider phenomenology, as in the case of a gravitino superWIMP, which requires large masses $\gae 500 \GeV$ for the NLSP of the MSSM. However, the mass of the dark matter candidate in these models is not fixed by the model, in which case the dark matter is not necessarily warm. The unique feature of the model we will present here is that the dark matter candidate is fixed by its relic density to be {\it necessarily warm}.

In the case of the MSSM, the only possible WDM candidate is a stable gravitino \cite{
wdmgravitino}. (Sterile neutrinos could also play this role in the MSSM extended to include neutrino masses.) However, this is not compatible with the MSSM in the case of gravity-mediated 
SUSY breaking as the gravitino is too heavy. The form of SUSY breaking may be determined 
at the LHC from the pattern of SUSY particle masses. 
Here we present a new WDM candidate, {\it the $Z_{2}$-singlino}, in a portal-like 
extension of the minimal supersymmetric (SUSY) standard model (MSSM)~\cite{portal_model,
portal_model2} in which the stability of the WDM particle is ensured by a $Z_2$-parity. The 
$Z_2$-singlino was previously introduced to provide 
a stable dark matter candidate in R-parity 
violating SUSY models \cite{portal_model}. Here we will show that, in a different region of 
its parameter space, the model 
provides a dark matter particle which is necessarily warm and with the right properties to explain non-singular galaxy halos and to alleviate the galaxy substructure and low angular momentum problems of CDM models.

{\bf The Model:} 
We extend the MSSM by adding a chiral superfield $\chi$ and messenger field $S$ of 
mass $M_S$. We also impose an additional $Z_2$ symmetry under which $\chi$ is odd, while 
all other fields are even. The effective superpotential after 
integrating out $S$ is given by~\cite{portal_model}
\begin{equation}
W= W_{\rm MSSM}+\frac{f \chi^2 H_u H_d}{M_S} + \frac{M_{\ochi_{0}}\chi^2}{2} .
\label{super_pot}
\end{equation}
Since $\chi$ is odd under $Z_2$, its lightest component cannot decay to any of the MSSM fields. 
Therefore the fermionic component of $\chi$, the $Z_2$-singlino, $\overline{\chi}$, is a 
good candidate for DM\footnote{The scalar partner of $\overline{\chi}$ is expected to gain a large mass from SUSY breaking.}. In Eqn. (\ref{super_pot}) 
we have included a SUSY mass $M_{\ochi_{0}}$ for $\chi$. Since in general the $\ochi$ mass 
must be small relative to the weak scale in order to account for WDM, a particularly 
interesting case is where the $\ochi$ mass is entirely due to the Higgs expectation values, 
with $M_{\ochi_{o}}$ zero or negligibly small. The small $\ochi$ mass can then be 
understood in terms of a large value for $M_{S}$ compared with the Higgs expectation 
values. The absence of a $\chi$ mass term in Eqn.(\ref{super_pot}) is 
guaranteed if there is an unbroken R-symmetry which allows the $\mu H_{u}H_{d}$ term 
of the MSSM, since in this case the R-charge of $\chi$ must be zero. 

$Z_2$-singlino dark matter is interesting as a SUSY implementation of gauge singlet 
dark matter. Gauge singlet scalar dark matter interacting via the Higgs 
portal~\cite{portals} was first proposed in~\cite{singlet_scalar} for the case of 
complex scalars and in~\cite{real} for real scalars, and was further discussed 
in~\cite{real2,real3,scalar_darkmatter1,scalar_darkmatter2,scalar_darkmatter3}. 
Couplings to hidden sector particles are currently of considerable 
interest~\cite{jasonkumar,hidden1,hidden2,hidden3,hidden4,doublet,seto,doublet2}. 
The simplicity of the terms in Eqn.(\ref{super_pot}) should allow them to easily form part of a hidden sector dark matter model.

In the following we will calculate the relic abundance of $\overline{\chi}$ as a function 
of its mass and reheating temperature, $T_{R}$.  We will show that in the case where the 
$\ochi$ mass is mostly due to the Higgs expectation value, $\ochi$ is {\it necessarily warm}
with a mass in the keV range
when it accounts for the observed dark matter density. 
For reasonable values of $T_{R}$ the primordial phase-space density can then 
account for that observed in dwarf spheroidal galaxies (dSphs), 
while the free-streaming length can damp the density perturbation on small scales
and so reduce galaxy substructure formation and angular momentum loss.

{\bf Relic Abundance of $\overline{\chi}$:} 
Production of $\overline{\chi}$ will occur mainly through thermal Higgs 
annihilations\footnote{ $\ochi$ can also be produced by decay of thermal Higgs particles. 
However, we find that this is negligible compared with Higgs annihilation except for $T_{R} 
\lae 300 \GeV$, where both processes give a similar contribution to $n_{\ochi}$. We will 
therefore neglect the contribution of Higgs decays in the following.}. We will see that 
most of the $\overline{\chi}$ are produced at temperatures close to the reheating 
temperature, $T_{R}$. In this case we can consider the Higgs expectation values to be 
zero and calculate with the weak eigenstate Higgs doublets. $\overline{\chi}$ production 
via thermal Higgs boson pair annihilation occurs due to the contact interaction in the 
Lagrangian $f \overline{\chi}\overline{\chi}H_{u}H_{d}/M_{S}$. The total rate of 
$\overline{\chi}$ production per Higgs pair annihilation is then 
\begin{equation}   \frac{d n_{\ochi}}{dt} + 3Hn_{\ochi} = \Gamma_{\ochi} n_{H} \;\;;\;\; 
\Gamma_{\ochi} = 8 n_{H}\sigma_{H} v_{rel}  ~,
\label{rateeq}
\end{equation}
In this $n_{H} = 2.4 T^{3}/\pi^2$ is the number density of a complex Higgs scalar, 
$\sigma_{H} = (f/M_{S})^{2}/16 \pi$ is the cross-section for  $h_{u}^{o}h_{d}^{o} 
\rightarrow \overline{\chi}\overline{\chi}$ (with the same for  $h_{u}^{+}h_{d}^{-} 
\rightarrow \overline{\chi}\overline{\chi}$) and $v_{rel} = 2$ is the relative Higgs 
velocity. There is an overall factor of 2 in $\Gamma_{\ochi}$ since each Higgs pair 
annihilation produces two $\overline{\chi}$ particles. In addition, there will be 
$\overline{\chi}$ production via Higgsino-Higgs boson annihilation to $\overline{\chi} 
\chi^{0}$. This will increase the $\overline{\chi}$ production rate by approximately 
a factor of 2, which we have included in $\Gamma_{\ochi}$. In terms of $T$ we obtain
\begin{equation} \Gamma_{\ochi} =  \frac{2.4 M_{\ochi\;sb}^{2} T^3}{\pi^3 v^4 \sin^{2} 2 \beta}   ~,
\label{rate2}
\end{equation}
where $\tan \beta = v_{u}/v_{d}$ with $v = \sqrt{v_{u}^2 + v_{d}^{2}}=174 \GeV$.
$M_{\ochi\;sb}$ is the contribution to the $\ochi$ mass from electroweak symmetry breaking,
$ M_{\ochi\;sb} = f v^{2} \sin 2\beta/M_{S}   ~.$
This relates the symmetry breaking contribution to the $\ochi$ mass to the strength of 
the $\ochi$ interaction with the MSSM Higgs, $f/M_{S}$. 
As a result, the $\ochi$ relic density fixes the $\ochi$ mass in the case where
it is mostly due to symmetry breaking.   
The total $\ochi$ mass is then 
$M_{\ochi} = M_{\ochi_{0}} + M_{\ochi\;sb}$. Since $\Gamma_{\ochi} \propto T^3$ while 
$H \propto T^2$ during radiation-domination, the production of $\chi$ will occur mostly 
at the highest temperature during radiation-domination, which is the reheating temperature, 
$T_{R}$. (The temperature can be higher during the inflaton-dominated era before reheating, 
but since $H \propto T^4$ during this era, its contribution to $\ochi$ production is small 
compared with production at reheating.) We define the thermalization temperature
 $T_{th}$ by the condition $\Gamma_{\ochi} = H \equiv k_{T}T^{2}/M_{Pl}$, where $k_{T} = 
(4 \pi^{3}g(T)/45)^{1/2}$ and $g(T)$ is the number of relativistic degrees of freedom. 
Therefore
\begin{equation}
 T_{th} = \frac{k_{T} \pi^{3} v^{4} \sin^{2}2\beta}{2.4 M_{\ochi\;sb}^{2} M_{Pl}}     ~.
\label{th}
\end{equation}
Assuming $g(T)$ is constant, Eqn. (\ref{rateeq}) can be written as 
\begin{equation} \frac{d}{dT}\left(\frac{n_{\ochi}}{T^3}\right) = -\frac{1}{T_{th}} 
\frac{n_{H}}{T^{3}}    ~, \end{equation} 
where we have used the relation $\Gamma_{\ochi}/HT = 1/T_{th}$, which is generally true 
during radiation-domination. Integrating this from $T_{R}$ to $T$, and noting that 
$n_{H}/T^{3} = 2.4/\pi^{2}$ is a constant, we obtain
\begin{equation} \frac{n_{\ochi}}{T^3}  =  \frac{2.4 (T_{R}-T)}{\pi^{2} T_{th}}          ~. 
\end{equation}
Clearly most of the production of the comoving $\ochi$ density occurs at $T \approx T_{R}$. 
Including an entropy dilution factor for the change of $g(T)$ from $T_{R}$ to the present 
CMB temperature, $T_{\gamma} = 2.4 \times 10^{-13} \GeV$, we find for the present $\ochi$ 
number density, 
\begin{equation} n_{\ochi}(T_{\gamma})  = \frac{2.4T_{\gamma}^{3}}{\pi^{2}} 
\frac{g(T_{\gamma})}{g(T_{R})} \frac{T_{R}}{T_{th}}   ~. \end{equation}
Using Eqn.(\ref{th}) to eliminate $T_{th}$, the relic abundance of $\ochi$ from Higgs 
annihilations is then given by 
\begin{equation} \Omega_{\ochi} = \frac{4(1.2)^2}{\pi^{5}} \left(\frac{M_{\ochi\;sb}}
{M_{\ochi}}\right)^{2} \frac{g(T_{\gamma})}{g(T_{R})}  \frac{ T_{R} T_{\gamma}^{3}  }
{k_{T} v^4 \sin^{2} 2 \beta} \frac{M_{\ochi}^3 M_{Pl}}{\rho_{c}}
     ~, 
\label{omeg}
\end{equation} 
where $\rho_{c} = 8.1 \times 10^{-47}h^2 \GeV^4$ is the critical density. 
Therefore 
$$ M_{\ochi} = 1.92 \left(\frac{M_{\ochi}}{M_{\ochi\;sb}}\right)^{2/3} 
\left(\frac{\Omega_{\chi}h^{2}}{0.113}\right)^{1/3} ~$$
\begin{equation}     
\times \left(\frac{1 \TeV}{T_{R}}\right)^{1/3} \sin^{2/3}2 \beta  \keV ~,
\label{chimass}
\end{equation}
where we have used $g(T_{\gamma}) = 2$ and $g(T_{R}) = 228.75$, corresponding to the 
MSSM degrees of freedom, and we have expressed $\Omega_{\ochi}$ relative the observed 
dark matter abundance, $\Omega_{\ochi}h^2  = 0.1131 \pm  0.0034$ \cite{wmap}.

From Eqn. (\ref{chimass}) we see that in the case where the $\ochi$ mass is due to
symmetry breaking, $M_{\ochi}$ is automatically close to a keV. As a result, 
the dark matter in this model is {\it necessarily warm}.

\begin{figure}
\begin{center}
\epsfig{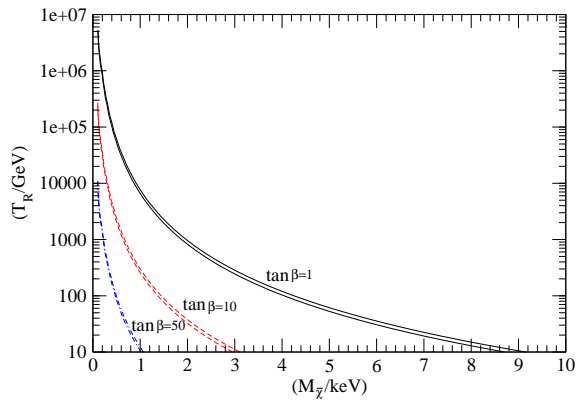}
\caption{Contours of $\Omega_{\ochi}h^2 = 0.1131 \pm  0.0034$ are shown in the plane of 
$M_{\ochi}$ versus $T_{R}$ for different values of $\tan \beta$. The $\ochi$ mass is set to 
its value from symmetry breaking, $M_{\ochi} = M_{\ochi\;sb}$.}
\label{fig}
\end{center}
\end{figure}

The relationship between $M_{\ochi}$ and $T_{R}$ for different values of  $\tan \beta$ 
is shown in Fig.\ref{fig} for the case $M_{\ochi\;sb} = M_{\ochi}$. From this we find 
that the range of $\ochi$ mass is tightly constrained. $T_{R} \lae 10^{6}$GeV is necessary 
in order to avoid thermal gravitino overproduction (for the case with hadronic gravitino 
decay modes) \cite{gravitino} while $T_{R} \gae 10^{2} \GeV$ in order to have the thermal 
Higgs necessary to produce the $\ochi$ density. In the case where the $\ochi$ mass is due 
to spontaneous symmetry breaking, $M_{\ochi\;sb} = M_{\ochi}$, the $\ochi$ mass is in the 
range 
\begin{equation}   0.19\; \sin^{2/3} 2 \beta \; \keV \lae M_{\ochi} \lae 4.1 \; \sin^{2/3} 
2 \beta \; \keV  ~. \end{equation}
Moreover, $T_R < T_{\rm th}$ is necessary in order to keep $\ochi$ out of thermal equilibrium, 
which is assumed in Eqn.(\ref{rateeq}). 
This gives $M_{\ochi}\gsim 0.3 \keV$. 
Thus for reasonable values of $\sin 2 \beta$ we 
expect $0.3 \keV \lae M_{\ochi} \lae 4 \keV$. 
The corresponding range of reheat 
temperature is $10^2 \GeV \lsim T_R \lsim 10^5 \GeV$. 
  
  In this we have used the observed abundance of dark matter as 
an input, which then determines the mass of the dark matter particle to be of the order of a keV. In doing so, we have implicitly tuned the 
mass of the dark matter particle for a given $T_{R}$. To put this tuning in perspective, we can compare the present model with the most popular dark matter scenario, that of thermal relic weakly interacting dark matter. In this case, the dark matter density is broadly of the correct order of magnitude when the dark matter particle mass and its interaction strength are determined by the weak scale. However, although phenomenologically encouraging, theoretically this amounts to an implicit tuning of the physics of thermal freeze-out against the completely unrelated physics of baryogenesis in order to obtain abundances of baryons and dark matter which are within a factor of 6. In other words, the interaction strength, which is determined by its mass scale, must be tuned in order to obtain the correct dark matter abundance relative to the baryon density. Similarly, in our model we are tuning the interaction strength, which is now determined by the dark matter particle mass, to obtain the correct dark matter abundance relative to the baryon density. In the thermal relic case the output is an interaction mass scale of the order of the weak scale. In our case the output is a mass for the dark matter particle of the order of a keV. Therefore, from a theoretical point of view, the two models are not dissimiliar in their need for tuning. Such tuning is a generic problem for any dark matter model which does not directly address the baryon-to-dark matter ratio.


{\bf The Phase-Space Density of $\overline{\chi}$:} 
Dark matter with a finite phase-space density may be able to explain the 
finite density of galaxy cores \cite{dal}. The coarse-grained phase-space density is defined by 
$Q \equiv \rho_{\overline{\chi}}/\sigma_{\overline{\chi}}^3 $, where the 1-D velocity dispersion 
is $\sigma_{\overline{\chi}}=\sqrt{(1/3)\langle \vec{P}_{\overline{\chi}}^2/M_{\overline{\chi}}^2 \rangle}$ 
\cite{dal}. If $Q$ is finite then there is a limit on how large $\rho$ can be for a given velocity 
dispersion, which prevents the formation of singular galaxy cores. $Q$ can be expressed in terms of 
the distribution of $\ochi$ produced by thermal Higgs annihilation ~\cite{devegaetal}
\begin{equation}
Q \equiv  \frac{ 3^{3/2}M_{\overline{\chi}}^3 \rho_{\overline{\chi}}}{\langle 
\vec{P}_{\overline{\chi}}^2\rangle^{3/2}}=\frac{3^{3/2} M_{\ochi}^{3} \Omega_{\overline{\chi}}
\rho_{c}J^{3/2}}{T_{d}^3}  \,,
\label{phase-space}
\end{equation} 
where $T_{d} = (g(T_{\gamma})/g(T_{R}))^{1/3} T_{\gamma}$ for a decoupled 
density created at $T_{R}$ and $ J = \int_0^\infty y^2 f(y) dy/\int_0^\infty 
y^4 f(y) dy $. Here $f(y)$ is the 
distribution function for production of $\overline{\chi}$ from thermal Higgs annihilation, 
where $y=|\vec{P}_{\overline{\chi}}|/T_{d}$ is the comoving momentum. Therefore 
\begin{equation}
Q =  5.6 \times 10^{7}
\left(\frac{\Omega_{\ochi} h^2}{0.113}\right) 
\left(\frac{M_{\ochi}}{1 \keV} \right)^{3}
 \left(\frac{J}{0.1}\right)^{3/2} \frac{{\rm eV}/{\rm cm}^3}{({\rm km/s})^3}~.   
\label{qmass} 
\end{equation} 
The distribution function due to thermal Higgs annihilation is not known at present. 
However, $J$ may be estimated by assuming that the momentum of the $\ochi$ at $T_{R}$ is equal to 
the mean momentum of the thermal Higgs particles, $p \approx 3 T$, in which case $f(y) = 
\delta(y - 3)$. This gives $J = 1/y^2 = 1/3^{2} = 0.11$. This is in good agreement with the 
value obtained in the case of thermal Higgs decay, $J = 0.12$~\cite{sterile_neutrino}.

 $Q$ is conserved for an adiabatically contracting collisionless gravitational system \cite{dal}. 
More generally, $Q$ can only decrease from its primordial value during relaxation in collisionless 
systems. Simulations of structure formation have shown that the phase-space density can decrease 
by a factor $10^2$ to $10^3$ \cite{phasedec}. 

      If the distribution of dark matter in dwarf spheroidal galaxies 
(dSphs) is cored, then a synthesis of recent photometric and kinematic dSph data \cite{gilmore}, which indicates a mean density $\sim 
5 {\rm GeV/cm^3}$ and central velocity dispersion $\sim 10 {\rm km/s}$, implies that~\cite{devegaetal}
\begin{equation} Q_{{\rm dSph}} \approx 5\times 10^6\;
\frac{{\rm eV}/{\rm cm}^3}{({\rm km}/{\rm s})^3}\,.
\label{dsph}
\end{equation} 
Eqn. (\ref{qmass}) then provides an upper limit on $M_{\ochi}$ in the case with an explicit SUSY 
$\ochi$ mass. If we assume a dynamical suppression of $Q$ by at most a factor $10^{3}$, then the 
largest value of $Q$ consistent with observations is $\approx 10^{10} {\rm eV}/{\rm cm}^3/
({\rm km}/{\rm s})^3$. From Eqn.(\ref{qmass}) the largest mass compatible with this is 
$M_{\ochi} \approx 6 \keV$.  This is not much larger than the mass range 0.3-4 keV implied by the 
relic density in the case with $M_{\ochi} \approx M_{\ochi\;sb}$. Since there is no reason 
to expect a SUSY mass for $\ochi$ in the keV range, it is more natural to assume that 
$M_{\ochi_{0}}$ is zero or negligible, with the small $\ochi$ mass then generated by electroweak 
symmetry breaking together with $M_{S} \gg v$.

Using Eqn. (\ref{chimass}) to eliminate $M_{\overline{\chi}}$ from Eqn.(\ref{qmass}), the phase 
space density of $\overline{\chi}$ as a function of $T_{R}$ is given by  
$$ Q = 4.0 \times 10^{8} \; \sin^{2} 2 \beta \left( \frac{M_{\ochi}}{M_{\ochi\;sb}} \right)^{2} $$ 
\begin{equation} \times \left(\frac{\Omega_{\ochi}h^{2}}{0.113}\right)^{2} 
\left( \frac{1 \TeV}{T_{R}} \right) 
 \left(\frac{J}{0.1}\right)^{3/2}  \frac{{\rm eV}/{\rm cm}^3}{({\rm km/s})^3}   ~. 
\label{Q}
\end{equation}  
Comparing with Eqn. (\ref{dsph}), we see that the value of $Q$ can be equal to or 
larger than $Q_{{\rm dSph}}$ for reasonable values of $T_{R}$. 
From Eqn. (\ref{Q}) the required reheating temperature is 
\begin{equation} T_{R} \approx 80 \; \sin^{2} 2 \beta \left(\frac{Q_{{\rm dSph}}}{Q}\right) \left( \frac{M_{\ochi}}{M_{\ochi\;sb}} \right)^{2}     \TeV   ~.
\label{qs}
\end{equation} 
Thus in the case where the $\ochi$ mass is mostly from the Higgs expectation value and the phase-space density of dSph corresponds to the primordial phase-space density without suppression, 
$10 \TeV \lae T_{R} \lae 100 \TeV$ is typically required, depending on $\sin 2 \beta$.  From 
Eqn.(\ref{chimass}) the corresponding $\ochi$ mass is $M_{\ochi} \approx 0.45 \keV$.  
If the primordial phase space density is suppressed during formation of dSphs, then larger $Q$ and lower $T_{R}$
are required.  In general, for $\sin 2\beta \gae 0.1$ and $10^{2} \GeV \lae T_{R} 
\lae 10^{5} \GeV$, $Q$ due to $\ochi$ dark matter is in the range $10^{5}-10^{10}\; {\rm eV}/{\rm cm}^3/({\rm km}/{\rm s})^3$.

{\bf Free-streaming length of $\overline{\chi}$:}
The free-streaming length, $\lambda_{\rm fs}$, below which primordial perturbations are suppressed, is roughly equal to the horizon when the $\ochi$ particles become non-relativistic. 
In general, for a distribution of relativistic decoupled particles, $\lambda_{fs}$ is given by \cite{fs}
\begin{equation}   \lambda_{fs} \approx 1.2 \; {\rm Mpc} \left( \frac{1 \keV}{M_{\ochi}}\right)
 \left(\frac{10.75}{g(T_{R})}\right)^{1/3} \left(
\frac{\langle p/T \rangle}{3.15} \right)    ~,
\label{fs}
\end{equation}
where  $\langle p/T \rangle$ is the mean momentum over $T$ of the initial relativistic $\ochi$ distribution. Since the 
$\ochi$ are produced by annihilation of thermal Higgs at $T \approx T_{R}$, we expect that their mean 
momentum will be approximately equal that of the thermal Higgs at $T_{R}$, such that $\langle p/T \rangle \approx 3$.
Thus with $g(T_{R}) =228.75$ and $0.3 \keV \lae M_{\ochi} \lae 4 \keV$, $\lambda_{fs}$ is in the range 
0.3 - 4 Mpc. (Smaller values are possible if the mean momentum of the $\ochi$ distribution is less than thermal.) 
The lower end of this range is comparable with the scale of galaxies and so may
alleviate the problems of overproduction of substructure and low angular momentum observed in CDM simulations \cite{satellite_problem,angular_momentum}.  

{\bf Lyman-$\alpha$ constraints:}
 Observation of Lyman-$\alpha$ 
absorption spectra constrains the matter power spectrum on small-scales and so provides a lower bound on the WDM particle mass given its momentum distribution \cite{ly1}.  Lower bounds on the mass of sterile neutrino WDM were obtained in \cite{ly2,ly3,lya}. In \cite{lyb}, lower bounds were obtained for WDM momemtum distribution functions which are generalizations of the Fermi-Dirac distribution. In the case of thermal relics which decoupled while relativistic, a lower bound of 1.7 keV (95$\%$ c.l.) 
was obtained, while for non-resonantly produced sterile neutrinos (less-than-equilibrium density but with a thermal momentum distribution) the corresponding lower bound was 9.5 keV \cite{lyb}. 

      Since the distribution function from thermal Higgs annihilation is not expected to be of Fermi-Dirac form, we cannot directly apply the results of existing Lyman-$\alpha$ analyses\footnote{Similarly, constraints on the WDM particle mass from globular cluster-based observations of the phase space density of the Fornax dwarf galaxy 
\cite{strigari} cannot be directly applied to the present model.}. 
However, as the $\ochi$ density from thermal Higgs annihilation is less than the thermal equilibrium density but the mean $\ochi$ momentum is of the order of the thermal Higgs momentum at $T \approx m_{H}$, we expect the Lyman-$\alpha$ lower bound on the $\ochi$ to lie between the lower bounds of \cite{lyb}, which may allow a window below the 4 keV upper bound from the $\overline{\chi}$ relic density. 

     In addition, Lyman-$\alpha$ observations can easily be consistent with 
WDM in the case where there is a significant ($\gae 40 \%$) component of CDM \cite{lyb}. Since the neutralino is a CDM candidate in our model in the case where 
there is an unbroken R-parity, mixed dark matter is a possibility. This does not diminish the advantage of a dark matter candidate which is necessarily warm.             

{\bf Discussion and Conclusions:} We have shown that the $Z_{2}$-singlino can account for the observed phase-space density of dwarf spheroidal galaxies, which may be evidence of non-singular cores.
In the case where the $\ochi$ mass comes entirely from the Higgs expectation value, 
$M_{\ochi}$ is fixed by the $\ochi$ relic density.  
The observed abundance of dark matter implies that $M_{\ochi}$ is in the range 0.3-4 keV, which coincides 
exactly with the range required for $\ochi$ to act as WDM. 
Therefore dark matter is {\it necessarily warm} in this model. 
The model can directly account for the phase-space density 
of dwarf spheroidal galaxies when $T_{R} \approx 10-100 \TeV$, while dynamical suppression of the primordial phase-space density allows smaller values of $T_{R}$ to be consistent with dSphs. The free-streaming length is in the range 
0.3-4 Mpc, the lower end of which may reduce the overproduction of satellites and loss of angular momentum observed in CDM simulations of galaxy formation. The small mass of $\ochi$ can be understood in terms of a large messenger mass, $M_{S} 
\approx 10^{10} \GeV$.  Such a heavy $S$ field might be identified with the messenger sector of gauge mediated 
SUSY breaking models. We will return to this possibility in a future study~\cite{mands}.

    Depending on $T_{R}$ and $\sin 2 \beta$, a range of primordial phase-space densities can be generated, with 
$Q \approx 10^{5}-10^{10} \; {\rm eV}/{\rm cm}^3/({\rm km}/{\rm s})^3$ when  $\sin 2\beta \gae 0.1$ and $10^{2} \GeV 
\lae T_{R} \lae 10^{5} \GeV$. This may allow the wide range of observed values of $Q$, ranging from order $10^{6} \; 
{\rm eV}/{\rm cm}^3/({\rm km}/{\rm s})^3$ in dSphs to $10^{4} \; {\rm eV}/{\rm cm}^3/({\rm km}/{\rm s})^3$ or less 
in normal spiral galaxies \cite{dal2}, to be understood, for example by having the values of $Q$ in dSphs close to 
their primordial values and the values in normal spirals suppressed by non-adiabatic evolution during 
structure formation. It has been suggested that too many dSphs may be generated when the mean 
primordial $Q$ is equal to that in dSphs \cite{madsen,wdmgravitino}. In that case a possible solution might be to have 
a mean primordial $Q$ much smaller than that observed in dSphs, with dSphs then forming from a high phase-space 
fraction of the $\ochi$ particles in the low momentum tail of the distribution \cite{madsen,wdmgravitino}. 

      In addition, strong constraints on $\ochi$ dark matter may be expected from Lyman-$\alpha$ observations. 
However, existing constraints on WDM masses cannot be directly applied as the $\ochi$ 
momentum distribution function due to thermal Higgs annihilation differs from those considered in existing studies.  
We will return to these issues in a future study~\cite{mands}.  

   If $R$-parity is unbroken in the MSSM then the model can be extended to a mixed 
dark matter model, with the $R$-stabilized MSSM LSP providing CDM in addition to 
the $Z_{2}$-stabilized $\overline{\chi}$ WDM. In this case  
WDM should easily be compatible with Lyman-$\alpha$ constraints.  

      Testing the model at colliders will be challenging due to the small effective coupling 
of $\ochi$ to the Higgs bosons, with the $h \ochi \ochi$ coupling being of the order of $v/M_{s} \approx 10^{-8}$. However, the simple form of the superpotential may allow the model to form part of a more complete model which could have distinctive collider signatures. 

{\bf Acknowledgement:}
JM and NS were supported by the European Union through the Marie Curie Research and 
Training Network "UniverseNet" (MRTN-CT-2006-035863).

\end{document}